\documentclass[journal]{IEEEtran}
\usepackage{soul}
\usepackage{epsfig}
\usepackage{cite}
\usepackage{url}
\usepackage{amssymb}
\usepackage{amsmath}
\usepackage{multirow}
\usepackage{enumerate}
\usepackage{float}
\usepackage{epstopdf}
\usepackage{bm}
\usepackage{amsmath}
\usepackage{amsthm}
\usepackage{amssymb}
\usepackage{subfigure}
\usepackage{algorithmic}
\usepackage{algorithm}
\usepackage{array}
\usepackage{color}

\def\linespaces{1}
\def\baselinestretch{\linespaces}

\newtheorem{lemma}{Lemma}

\newcommand{\warn}[1]{}
\DeclareMathSizes{10}{9}{7}{5}

\begin{document}

\title{Energy-efficient Analytics for Geographically Distributed Big Data}

\author{Peng Zhao, Shusen Yang, Xinyu Yang, Wei Yu, and Jie Lin
\thanks{Peng~Zhao, Xinyu~Yang, and Jie Lin are with the Department of Computer Science and Technology, Xi'an Jiaotong University, Xi'an 710049, China. Emails: \{p.zhao, yxyphd, jielin\}@mail.xjtu.edu.cn.}
\thanks{Wei~Yu is with the Towson University, MD 21252. E-mail: wyu@towson.edu.}
\thanks{Shusen~Yang is with the Institute of Information and System Science at Xi'an Jiaotong University, China; and the Department of Computing at Imperial College London, United Kingdom. Email: cuiangmu@gmail.com.}% <-this % stops a space
}

\maketitle

\begin{abstract}
Big data analytics on geographically distributed datasets (across data centers or clusters) has been attracting increasing interests in both academia and industry, posing significant complications for system and algorithm design. In this article, we systematically investigate the geo-distributed big-data analytics framework by analyzing the fine-grained paradigm and the key design principles. We present a dynamic global manager selection algorithm (GMSA) to minimize energy consumption cost by fully exploiting the system diversities in geography and variation over time. The algorithm makes real-time decisions based on the measurable system parameters through stochastic optimization methods, while achieving the performance balances between energy cost and latency. Extensive trace-driven simulations verify the effectiveness and efficiency of the proposed algorithm. We also highlight several potential research directions that remain open and require future elaborations in analyzing geo-distributed big data.
\end{abstract}

\IEEEpeerreviewmaketitle


\section{Introduction}\label{section1}

\begin{figure}
\centering
\includegraphics[height=6.8cm]{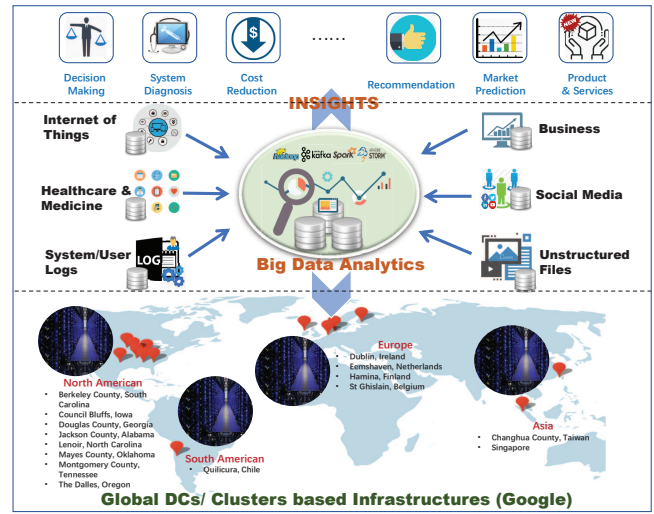}
\vspace{-2mm} \caption{Illustration of  big data systems and applications.} \label{bigdata_analytics}
\vspace{-2mm}
\end{figure}

The emerging ICTs (Information Communications Technologies) have led to a deluge of big data from a variety of domains, such as user-generated data, system logs, healthcare and scientific sensors, social networks, business companies, and supply chains systems. As illustrated in Figure~\ref{bigdata_analytics}, big data analytics \cite{lavalle2011big}  has demonstrated its great potential in capturing valuable insights for improving the decision making, system diagnosis, risk minimization, and develop new products and services. A new IDC (International Data Corporation) report\footnote{http://www.idc.com/} predicts that the big data and business analytics market will grow from \$130 billion by the end of this year to \$203 billion by 2020. That's a compound annual growth rate (CAGR) of 11.7\,\% over the next years.

Big data analytics extracts useful knowledge from huge digital dataset through advanced machine learning and data mining algorithms, which are normally computational-intensive and require intelligent, efficient, and scalable engines for deploying analysis services, programming tools, and applications. Generally, the data center (DC)-based computing infrastructure serves as an effective platform for satisfying both the computational and data storage requirements of big data analytics. To meet increasing data analysis demands and provide reliability, service providers deploy their data analytics service globally on multiple geographically distributed DCs, referred to as the Geographically-distributed Data Analytics (GDA) \cite{pu2015low, dolev2017survey}. The basic infrastructures for GDA generally consist of a massive number of servers and multiple Internet Data Centers (IDCs) in different locations. For instance, Google has deployed more than a dozen data centers across the U.S., South America, Europe, and Asia, with more than 2.5 million servers. Figure~\ref{bigdata_analytics} illustrates the geographical distribution of the DCs of Google. Compared with data analysis using a single DC, GDA adaptively performs distributed and parallel processing and analytics across multiple DCs, resulting in a much more scalable, fast-response, and resilient big data solution.

While significant benefits have been offered by GDA, megawatts of electricity are consequently required to power these extensive infrastructures, consisting of computational servers, storage devices, network equipments, cooling systems, and so on. As a result, millions of dollars must be spent on electricity by the infrastructure holder or service provider. According to the investigation \cite{datacenterenergy}, U.S. DCs consumed about 70 billion kilowatt-hours of electricity in 2014, representing representing 2\,\% of the country's total energy consumption and tremendous financial cost, that is equivalent to the amount consumed by about 6.4 million average American homes that year. Thus, a reduction of energy consumption by even a small percentage can result in considerable monetary savings. Furthermore, the resultant environment pollution is another serious consequence of huge energy consumption. The global carbon emissions from DCs accounted for approximately 0.6\,\% of the total in 2008, and this fraction is expected to reach 2.6\,\% by 2020 \cite{koomey2011growth}. For instance, the carbon emission of Google's DCs in the U.S. is approximately $1.13 M$ tons in 2010, which is equivalent to that emitted by 280,000 cars \cite{gao2012s}. Therefore, energy efficiency has played an enormous role in taming the data center industry.

There exist a large body of work on the issue of reducing energy consumption of DCs \cite{hameed2016survey}. The early intuitive approaches focus on the coping strategies within the DC through hardware optimization, dynamic capacity righg-sizing, virtual machine (VM) migration, and using renewable energy. A promising solution recently attempts to achieve energy efficiency through the dynamic job requests allocation across geo-distributed DCs by exploiting the geographic diversities in energy efficiency, electricity price, and power utilization. For example, the cost-aware schedule approach \cite{qureshi2009cutting}, which suggests to map the job requests to the DC with lower local electricity price, is an efficient way of reducing the electricity bill. Nonetheless, most of the state-of-the-art approaches cannot be directly applied to the applications of GDA. Due to the geo-distributed data placement, the job processing pattern of GDA is much different from the traditional applications in the cloud, where the job can be handled by the single DC with its local datasets independently. More specifically, the following two challenges should be further elaborated:

\begin{itemize}
  \item 1)	What are the impacts of data-intensive job allocation among geo-distributed DCs on the global energy consumption? To reduce the response time, the modern big data analytics service for each job generally requires the distributed, parallel task processing across multiple different data centers simultaneously. Selecting suitable global manager plays an important role in the global energy consumption.
  \item 2)	How to deal with system and data uncertainties such as the stochastic job arrivals, computing resource, energy price, and data availability for different DCs. Various uncertainties in the system have significantly impact on the benefits of the service provider (including the directly income from data analytic service and the expenditure from energy consumption) and the user experience (for example, the latency of data analytics). This is a tradeoff problem should be considered in future.
\end{itemize}

To address above two challenges, we have developed a realistic energy model for stochastically arriving GDA job. In addition, we have developed a novel energy-efficient algorithm to dynamically dispatch the job requests to geographically distributed DCs. The algorithm minimizes the cost from energy consumption while achieving a corresponding tradeoff in latency performance by maintaining the stable queue backlogs. The algorithm makes greedy job dispatching based on the measurable system parameters (for example, queue backlogs, energy price, and so on) without predicting the future knowledge of the stochastic system states, such as job arrival, distribution of datasets, and specific parallel precessing pattern of big-data analytics. The extensive experiments based on the real data traces of Facebook DCs demonstrate the effectiveness of our algorithm in terms of energy cost and delay performance.

\section{Big Data Analyitics over GD-DCs}\label{section2}

\begin{figure*}
\centering
\includegraphics[height=7.5cm]{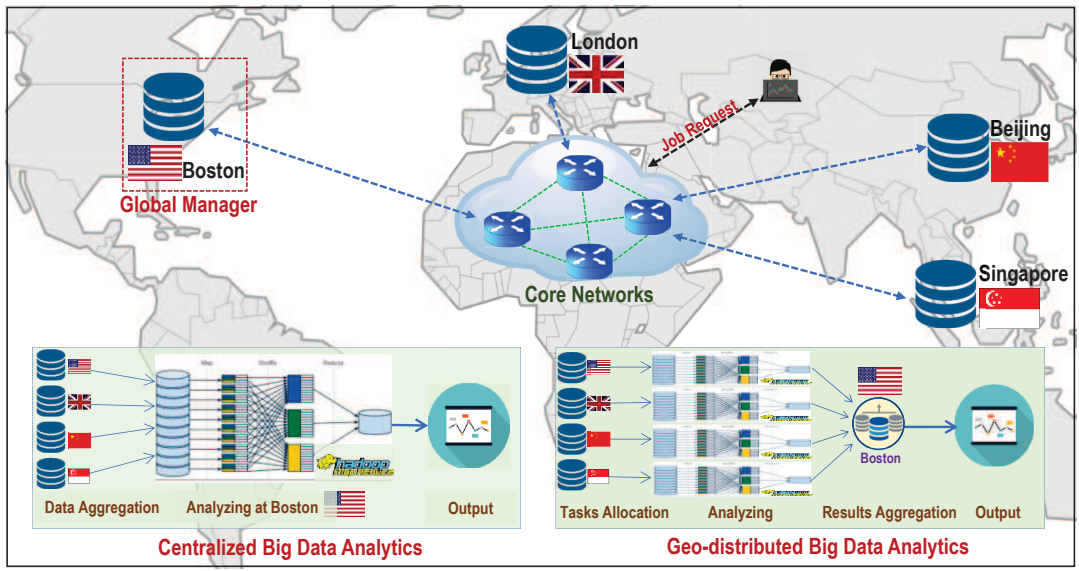}
\vspace{-2mm} \caption{Big-data analytics over geo-distributed datasets: Centralized Big Data Analytics vs. Geo-distributed Big Data Analytics.} \label{gdda}
\vspace{-2mm}
\end{figure*}

Currently, the widely-used approach adopted by many companies performs the geo-distributed data analytics by gathering the required datasets across all related sites to a central data center (for example, a more powerful one), where the job is processed with standard single-cluster technologies, such as Hadoop-based stacks \cite{vulimiri2015global}. As illustrated in Figure~\ref{gdda}, the DC in Boston (Global Manager) is responsible for handling job requests by gathering the datasets across four locations (Boston, London, Beijing, and Singapore), and then analyzing them centrally. Nonetheless, traditional big data analysis requires the centralized data aggregation, which becomes the major performance bottleneck of systems. For example, a well known Microsoft application for data analytics produces over 100 TB/day from multiple geo-distributed DCs into a centralized analytics stack \cite{kinshumann2011debugging}. More specifically, centralized big data aggregation results in the following issues. 

\begin{itemize}
  \item First of all, big-data aggregation across geo-distributed data centers significantly delays the timeliness of the analytics. For some real-time decision algorithms based on the results of data analytics, such latency is unacceptable \cite{yan2016tr}. Examples of time-sensitive data analytics include analyzing user logs to make advertisement push, analyzing the network traffic to detect the network attack (for example, DDoS).
  \item Second, the bandwidth cost of data transfer is another serious consequence that should be seriously considered \cite{greenberg2008cost,yassine2016bandwidth,laoutaris2011inter}. It is worth noting that, data transfer for big-data analytics has the characteristics of high frequency, which is much different from the traditional data transfer, such as bulk backup for load balance and fault tolerance.
  \item Moreover, considering the privacy preserving and regulatory concerns, it is unwise to transfer large volumes of original intensive data (for example, user activity logs) across multiple data centers, and the central data aggregation might be inadmissible in some practical systems \cite{mohan2012gupt,wu2016privacy}.
\end{itemize}

Clearly, the traditional centralized data analysis is not suitable or ideal for applications that span geo-distributed datasets. To address the above issues, the natural alternative approach is to execute analytics jobs in a geo-distributed manner through intra-DC analytics frameworks. Geo-Distributed Data Analytics (GDA) framework can be applied to most popular big data processing paradigms today. According to the type of input data, current big data analytics can be categorized into two alternative paradigms: \emph{Batch Processing} and \emph{Stream Processing}. Correspondingly, different analytics paradigms have different processing timeliness requirements.
\begin{itemize}
  \item \emph{Batch Processing}: In this paradigm, the data is first collected and stored, and then the data chunks are processed in parallel and distributed manner. Examples of input data for batch processing can include historical operational data, business data, social media data, service data, and so on. Batch processing has less timeliness requirement, for example, hours or perhaps even days. \emph{MapReduce} \cite{dean2008mapreduce} is a typical and prominent batch-processing model.
  \item \emph{Stream Processing}: The input data of this paradigm arrives as a real-time, continuous streams, such as click streams on web page, user request/query streams, message notifications, event monitoring streams. For instance, Google, Facebook and other advertising companies analyze the real-time user data (such as web browsing or search queries) to push specific advertisements. The streaming processing paradigm needs to complete each job in near-real-time -- probably seconds at most. Representative open source stream-processing systems include Apache Storm \footnote{http://storm.apache.org/} and Spark Streaming \footnote{http://spark.apache.org/}.
  %\item \emph{Interactive Query}:
\end{itemize}

In this article, we consider the geo-distributed big-data analytics framework to logically span all DCs, as illustrated in Figure~\ref{gdda}. The geo-distributed sites are globally connected using a core networks. Without losing generality, we assume that the bandwidth between the core network and specific site could be significantly heterogeneous due to widely different link capacities and other applications sharing the same links. The complete datasets for analytics are geo-distributed spanning all DCs, that is, each site only holds a data subset. Examples of such data could include the user logs of an application, the network logs of the subsystem, the application data uploaded by the users. The jobs of data analytics are launched by the user and arrive at the system randomly. It is worth noting that the job launcher can be any participator, such as the system decision maker and the service subscriber. The system then makes an important decision of selecting an appropriate site as the global manager. As the global manager, each data analytics job will be decomposed into a DAG of \emph{stages}, each of which consist of multiple parallel \emph{tasks}. These tasks will be allocated to geo-distributed sites for parallel processing according to the specific strategies, such as datasets placement, system capacity, latency, network I/O, and so on.

Taking the typical data analytics paradigm \emph{MapReduce} as an example. Traditionally, as the central \emph{MapReduce}, the input dataset for analytics first need to be gathered from geo-distributed data centers to a central one (we could also call this data center "global manager").  Then, the data analytics job is broken down to as many as \emph{Map} tasks as input blocks and one or more \emph{Reduce} tasks. Finally, these tasks are executed in parallel. Differently, geo-distributed \emph{MapReduce} dose not required the central aggregation of geo-distributed datasets. The global manager first breaks the job into multiple \emph{Map} and \emph{Reduce} tasks, then these tasks are allocated to different data centers and executed in parallel. The specific tasks allocation strategy can be quite different based on different considerations, including data placement, response time, bandwidth cost, and so on. For instance, study in \cite{pu2015low} present a system (\emph{Iridium}) for low latency geo-distributed analytics, which is achieved by optimizing placement of both data and tasks. It is worth noting that GDA does not mean that there is absolutely no data transfer between different data centers. For instance, some intermediate data of \emph{MapReduce} (for example, the output of \emph{Map} tasks) also need to be exchanged for the subsequent processing of \emph{Reduce} tasks.

In addition to the task allocation strategy, the global manager selection is clearly an important system decision for the overall performance of GDA. In particular, the issues of energy consumption, latency, geographic diversities, and the system uncertainties should be seriously considered in global manager selection. In this article, we study the problem of selecting the optimal global manager to minimize the cost of energy consumption by considering various diversities and uncertainties in the system, while providing the efficient data analytics service with respect to latency.

\begin{figure*}
\centering
\includegraphics[height=6.8cm]{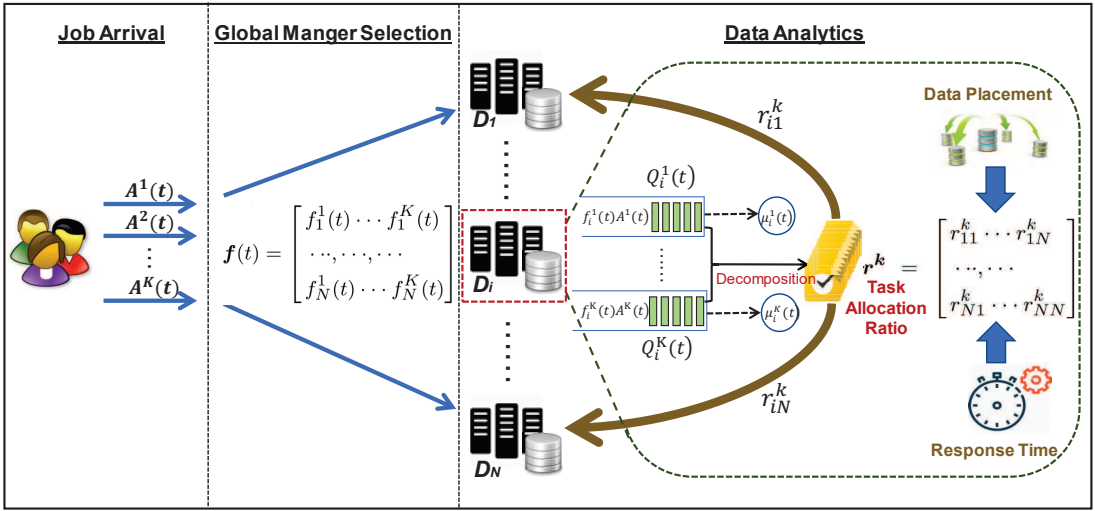}
\vspace{-2mm} \caption{Framework of geo-distributed data analytics} \label{model}
\vspace{-2mm}
\end{figure*}

\section{Energy Consumption of GDA}\label{section3}

In this section, the energy consumption of GDA is analyzed to capture the characteristics in geographic diversities, which is much different from the traditional central data analytics. In GDA, the energy consumption of each job comes from multiple data centers which performs the parallel tasks simultaneously. Generally speaking, the power consumption of a data center could be modeled as a function of the power consumption of Information Technology (IT) equipments (for example, servers, storage devices, and network equipment) and its corresponding Power Usage Effectiveness (PUE).

PUE is an industry accepted ratio for the measurement of effective usage of electrical power. The PUE can vary widely in real time among individual data centers depending on infrastructure equipment configurations and efficiencies, time of year, and local climate. According to the investigation, we know that the data centers across all U.S. uses nearly 50\% of its total power on infrastructures other than the IT equipment in average. Thus, the diversity of PUE is significantly critical and cannot be ignored in modelling the power consumption. In practical, the near-real-time PUE can be obtained in many practical system, including Google and Facebook. For instance, Google calculates its PUE every 30 seconds.

Although the tasks of each job in GDA are executed geographically in parallel, the total number of tasks (or computational tasks) are not reduced actually; and thus the computation resource from IT equipments is approximately fixed for each job. Therefore, we assume that the power consumption of IT equipments for each job can be considered as a fixed value regardless of where the tasks are executed. As the global manager, the ration of normalized number of tasks allocated to each dd or task allocation ratio can be conveniently measured based on its historical data or the specific task allocation algorithm. Then, considering the diversities of different DC in energy efficiency (for example, PUE), the total energy consumption of each job can be expressed as a weighted accumulating from different data centers based on the task allocation ratio.

Power consumption consequently incurs system cost. On the one hand, the service provider or
system owner will pay monetary cost to the electricity providers. Due to the different power generation profiles, multiple electricity price market is ubiquitous. For instance, the electricity price may vary on an hourly or 15-min basis in some regions of the U.S. On the other hand, from the perspective of electricity generation, power consumption is accompanied with carbon emission and environment pollution. Electrical energy in a region is typically generated with different fuel types, such as coal, oil, gas, nuclear, wind, solar, and so on. Generation by burning fossil fuels emits much more carbon than generation with renewable energy (nuclear, wind, solar, and so on).  Nonetheless, there is a significant difference among the fuel mix in different regions. Therefore, the price market of energy consumption can vary significantly depending on different system consideration.

\section{Case Study of Global Manager Selection}\label{section4}

As the case study of optimal global manager selection, we first introduce the fine-grained system models to capture the characteristics of GDA in geographic diversities and formulate the problem as a constrained optimization problem. Then, the dynamic algorithm based on the stochastic optimization is presented to solve the problem.

\subsection{System Model and Problem Formulation}

\subsubsection{System model}

As illustrated in Figure~\ref{model}, we consider the geo-distributed big-data analytics framework to logically span all DCs. The system consists of multiple geo-distributed DCs, denoted as $D_1, D_2,..., D_N$. All DCs are globally connected using a core network. Without losing generality, the bandwidth between the core network and specific sites could be significantly heterogeneous due to widely differing link capacities and other applications sharing the same links. There are $K$ types of data analytics jobs served by the system. In other words, there are $K$ types of big dataset in the system. For each type of big data, the complete datasets are distributed across all DCs. The parallel tasks of each job possibly be processed in any DC. The system operates in slotted time, with slots normalized to one unit. Without losing generality, we assume that the job inter-arrival times are time much shorter than the length of a time slot. We now consider the energy cost caused by geo-distributed big data analytics.

\textbf{\emph{Job arrival and global manager selection:}} The data analysis jobs are launched by the users and arrive randomly. The launcher can be any participator in the system, such as system
decision maker or service subscriber. Denote the number of arrived type-$k$ jobs ($k=1, 2, ..., K$) during time slot $t$ as $A^k(t)$ with the arrival rate $\lambda^k$. We assume that there always exists a $A_{MAX}^k$ that satisfies $A^k(t)\leq A_{Max}^k$ for all types of job across all time slots. For any arrived job request, the system first selects one DC as its ¡°Global Manager¡± for globally handling the procedure. For job requests with type $k$ during time slot $t$, we decide $f_i^k(t) \in[0,1]$, the fraction of jobs with type $k$ which selects data center $D_i$ ($\Sigma_i f_i^k(t)=1$) as the global manager. Thus, for any global manager DC $D_i$ during time slot $t$, the total number of arrived job with type $k$ is $f_i^k(t) A^k(t)$. For ease of description, all decision variables during time slot $t$ is defined as $\boldsymbol{f}(t)=\left[ \begin{aligned}
  & f_{1}^{1}(t) \cdot \cdot \cdot f_{1}^{K}(t) \\
 & \cdot \cdot \cdot ,\cdot \cdot \cdot ,\cdot \cdot \cdot \\
 & f_{N}^{1}(t) \cdot \cdot \cdot f_{N}^{K}(t) \\
\end{aligned} \right].$

\textbf{\emph{Service engine of GDA:}} The data analysis service in each global manager can be considered as a series of dynamic queues. Specifically, each global manager maintains the queues of unfinished jobs. In our system, we assume that the global manager can accurately measure the queue backlogs of the unfinished job. Denote $Q_i^k(t)$ as the backlog of type-$k$ job in the queue of data center $D_i$ at time slot $t$. For ease of description, all queue backlogs during time slot $t$ is denoted as a vector $\boldsymbol{Q(t)}$.Then, every time slot the queue $Q_i^k(t)$ changes according to the following queueing law:
\begin{equation} \label{queue_change}
Q_i^k(t+1) = \max [Q_i^k(t)+f_i^k(t) A^k (t)-\mu_i^k(t),0],
\end{equation}
where $Q_i^k(t+1)$ is the queue backlog of type $k$ job of $D_i$ at the beginning of next time slot $t+1$, $\mu_i^k(t)$ is a random variable and presents the number of job (also called service rate) of type $k$ served by the global manager $D_i$ during time slot $t$. Due to geo-distributed task
processing, the number of jobs served by the global manager each time slot (also called service rate) is a random variable, which is closely associated with computational capacity of each DC, dataset distribution, dynamic I/O constraints (e.g., up and down bandwidth), and the specific task allocation algorithm. Moreover, many practical systems are not specially designed for data analysis, and also run many core business applications. We assume that there always exists a $\mu_{Max}^k$ that satisfies $\mu_i^k(t)\leq \mu_{Max}^k$ for all types of job across all time slots. Different from traditional central applications, the service rate in GDA system is much associated with the computational capacity, dataset distribution, network I/O constraints (that is, bandwidth), and the specific task allocation strategy of GDA system.

\textbf{\emph{Energy consumption cost of GDA:}}
For the type-$k$ job, we assume its fixed energy consumption from IT component $P^k$. Considering the time varying PUE of each data center, we denote the PUE of data center $D_i$ during time slot $t$ as $PUE_i(t)$. For each type of job, we assume that the system maintains a global task allocation ration $\boldsymbol{r^k}=\left[ \begin{aligned}
  & r_{11}^{k} \cdot \cdot \cdot r_{1N}^{k} \\
 & \cdot \cdot \cdot ,\cdot \cdot \cdot\\
 & r_{N1}^{k} \cdot \cdot \cdot r_{NN}^{k} \\
\end{aligned} \right]$, where $r_{ij}^k$ represents the ratio of tasks allocated to data center $D_j$ when the global manager $D_i$ processing the job of type $k$. Thus, for any global manager $D_i$, the energy consumption of processing a type $k$ job can be calculated as $\sum_{j=1}^{N} PUE_j(t)  r_{ij}^k  P^k$. Moreover, considering the varying cost market due to the geographic diversity and variation over time, we characterize the energy price at the location of $D_i$ by a weighted function associated with $t$, denoted as $\omega_i(t)$. The specific definition of $\omega_i(t)$ can be determined based on different considerations of electricity price and environmental damage. Corresponding, the corresponding energy cost of global manager$D_i$ for processing a type $k$ job during time slot $t$ can be modeled as $\sum_{j=1}^{N} \omega_j(t) PUE_j(t)  r_{ij}^k  P^k$. Therefore, for all types of job arrival during time slot $t$, the total energy cost of all data centers, denoted as $Cost(t)$, can be finally calculated as $Cost(t) = \sum_{k=1}^{K} \sum_{i=1}^{N} \sum_{j=1}^{N} f_i^k(t) C^k(t) \omega_j(t) PUE_j(t)  r_{ij}^k P^k$.

\begin{figure*}
\centering
\includegraphics[height=8.5cm]{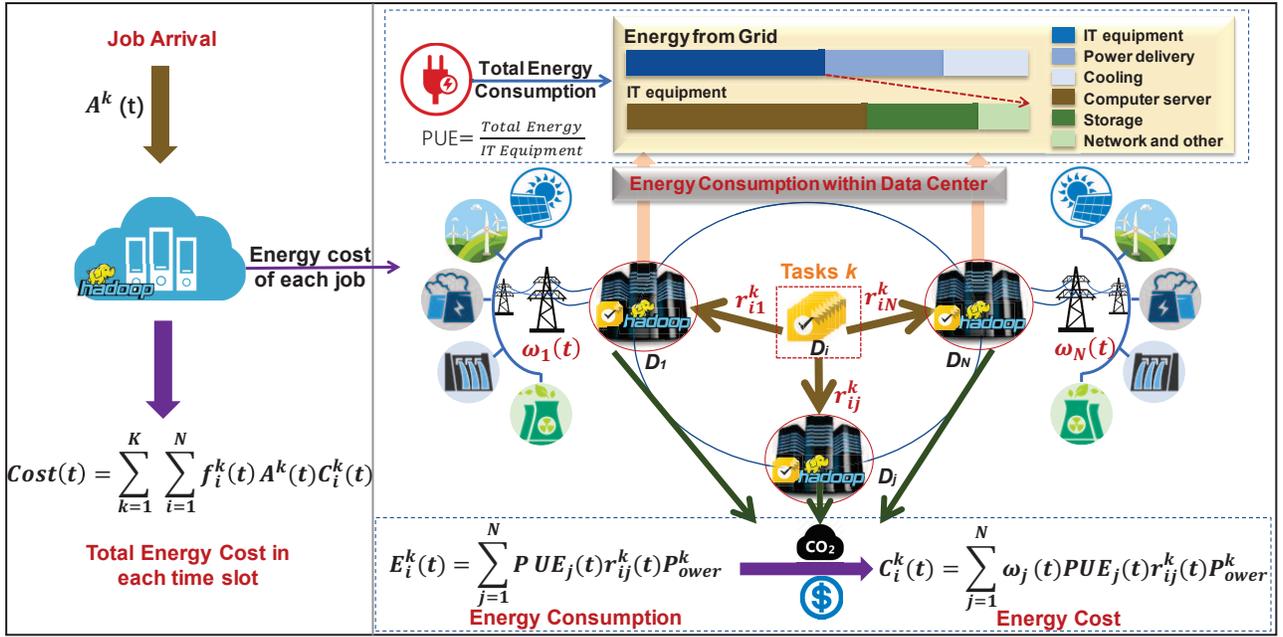}
\vspace{-2mm} \caption{Energy Consumption and Cost of each time slot in GDA} \label{energy}
\vspace{-2mm}
\end{figure*}

\subsubsection{Problem formulation}

In this article, we consider to minimize the energy cost for GDA by optimally selecting the appropriate global manager. Intuitively, it is beneficial to allocate the job to the data centers that can process the requests with the lowest cost as much as possible. Nonetheless, in a real-world system, energy cost is generally constrained by many other factors, such as the specific pattern of data analytics (for example, task allocation strategy), the geographic diversities in energy efficiency, and the delay requirement of the job. If a data center with limited processing efficiency is selected as the global manager, a long queue backlog may be incurred, increasing the response time, and potentially wearing out the users' patience.

Therefore, we consider the practical tradeoff between energy cost and delay performance in this article. To achieve the delay performance requirement, another desired system objective is maintaining the queue backlogs for system stability. Throughout this article, the strong system stability is defined as follows \cite{neely2005dynamic}:
\begin{equation} \label{stability}
\overline{Q} \triangleq \limsup_{t\rightarrow\infty} \frac{1}{t}  \sum_{\tau=0}^{t-1} \sum_{i=1}^{N} \sum_{j=1}^{K} \mathbb{E}\{Q_j^i(\tau)\} < \infty,
\end{equation}
where $\overline{Q}$ is the time average queue backlog of all data centers for all job types.

Therefore, we formulate the global manager selection to be an optimization problem, which minimizes the time average energy cost subject to the constraints of job arrival. Mathematically, we have the following constrained optimization problem:
\begin{eqnarray}\label{formulation}
\min_{\boldsymbol{f}(t)} & & \lim_{t\rightarrow\infty} \frac{1}{t}\sum_{\tau=1}^t Cost(\tau)  \\ \nonumber
\text{s.t.} & & \Sigma_i f_i^k(t)=1 \quad \forall k,  \\ \nonumber
                 & &  f_i^k(t) \geq 0 \quad \forall k \quad \forall i, \\ \nonumber
                 & &  and \quad (\ref{stability}).
\end{eqnarray}

Actually, the optimal global manager selection policy for optimization objective does not exist without the assumption of infinite delay requirement. In other words, the minimum energy cost can be achieved only in the system for delay tolerant jobs. In reality, there always exists an acceptable response time for users. Response time violation may consequently dent the system¡¯s appeal to clients, and thus reduce its competitiveness in the market. Thus, the practical control objective is to achieve a tradeoff between the minimum cost and delay.

\subsection{Dynamic Global Manager Selection Algorithm}

Considering the stochastic characteristic of the system in job arrival, energy efficiency, energy price, and service rate, the concept of Lyapunov optimization framework is used to solve the global manager selection problem for more practical control algorithm. The basic idea of our solution is to define a Lyapunov function that measures current queue backlogs first, and then makes greedy decisions to minimize the upper bound of Lyapunov function at every time slot. Such greedy decisions do not require the future knowledge of stochastic system states; and hence, offer a potential advantage in overcoming the complexity explosion problem.

More specifically, we first introduce a quadratic Lyapunov function for each time slot to measure the aggregate queue backlogs, denoted by $L(t)$. $L(t)$ is defined as $L(t) \buildrel \Delta \over = \frac{1}{2}\sum\limits_{i = 1}^N {\sum\limits_{k = 1}^K {Q_i^k{{(t)}^2}} }$. Then, the one-step conditional Lyapunov drift, denoted by $\Delta(t)$, is defined as the expected change in
the Lyapunov function over a single time slot, that is, $\Delta (t) \buildrel \Delta \over =\mathbb{E}\{ L(t + 1) - L(t)\left| \boldsymbol{Q(t)} \right.\}$. To stabilize the system while minimizing the energy cost, i,e,. the optimization objective in \ref{formulation}, the dynamic algorithm can be designed to make the greedy decision of selecting the global manager that minimizes the upper bound on a drift-plus-penalty function of each time slot, that is, $\Delta (t)+V Cost(t)$. The key derivation of the control algorithm is to obtain the upper bound of this function. We define the upper bound in the following lemma.

\begin{lemma}
For any feasible global manager selection strategy under the job allocation constraints, we have,
\begin{flalign} \label{lemma}
&\Delta(t) + \mathbb{E}\{V C(t)| \textbf{\emph{Q}}(t)\} &  \nonumber \\
&\leq B + E\{ \sum\limits_{i = 1}^N {\sum\limits_{k = 1}^K {Q_i^k(t)f_i^k(t){J^k}(t) + V Cost(t)} } \left| {\textbf{\emph{Q}}(t)} \right.\}&  \nonumber \\
 & \quad  - E\{ \sum\limits_{i = 1}^N {\sum\limits_{k = 1}^K {f_i^k(t){J^k}(t)\mu _i^k(t)} } \left| {\textbf{\emph{Q}}(t)} \right.\} &  \nonumber \\
  & \quad  - E\{ \sum\limits_{i = 1}^N {\sum\limits_{k = 1}^K {Q_i^k(t)\mu _i^k(t)} } \left| {\textbf{\emph{Q}}(t)} \right.\}&
\end{flalign}
Here, $B = \frac{1}{2}N\sum\limits_{k = 1}^K {J{{_{Max}^k}^2}}  + \frac{1}{2}N\sum\limits_{k = 1}^K {\mu {{_{Max}^k}^2}} $.
\end{lemma}

The detailed derivation and proof of Lemma 1 is similar to many work based on Lyapunov optimization theory \cite{neely2012stability} and thus is ignored here due to space limitation. According to the Lyapunov optimization theory, the strategy of solving the optimization problem requires to greedily minimize the right-hand-side of the inequality (\ref{lemma}) by controlling the global manager selection decisions $\boldsymbol{f}(t)$. More specifically, the above design strategy yields the following dynamic global manager selection algorithm (GMSA).

\emph{\underline{Global Manager Selection Algorithm(GMSA)}}: The following control operations are performed the each front-end server at every time slot $t$.

\begin{enumerate}
  \item \textbf{Job arrival and system states observation}. The system observes the current values of related system parameters and variables for each job type $k$ and DC $D_i$, including  job arrival $A^k(t)$, queue backlogs $Q_i^k(t)$, energy price weight $\omega_i(t)$, and the service rate $\mu_i^k(t)$.
  \item \textbf{Global manager selecting}. The global manager is selected (i.e., computing $\boldsymbol{f}(t)$ shown in Figure~\ref{model}) through solving the following  Linear Programme (LP) problem.

      \begin{eqnarray*} \label{objective}
      &&\textbf{Minimize}\quad \sum\limits_{i = 1}^N {\sum\limits_{k = 1}^K {(f_i^k(t){A^k}(t)(Q_i^k(t)}}-\mu _i^k(t))\\ \nonumber
      && \qquad\qquad\qquad \qquad-Q_i^k(t)\mu _i^k(t)) + VCost(t).\\
      && \rm \textbf{Subject to} \quad \Sigma_i f_i^k(t)=1 \quad \forall k,  \\ \nonumber
                 & &  \quad\qquad \qquad  f_i^k(t) \geq 0 \quad \forall k \quad \forall i, \\ \nonumber
                 & &  \quad\qquad \qquad  and \quad (\ref{stability})..
      \end{eqnarray*}
      where $V$ is a positive  parameter to control the performance tradeoff between the energy cost and queue backlogs. Correspondingly, the practical job allocation operation is then carried out to dispatch the jobs to different DCs as the global manager for data analytics.
  \item \textbf{Job processing and queue updating}. Each global manager handles the allocated  the jobs and performs the corresponding data analytics. Moreover, the queues of each DC are updated based on the newly arrived jobs and completed jobs during this time slot.
\end{enumerate}
\vspace{0.3em}

Notice that the decision made by \emph{GMSA} at each time slot only depends on the current measurable system parameters, and does not require knowledge of random future events, such as job arrival, service rate, weight price, and so on. All that is needed is for each back-end DC to broadcast their current queue backlogs and all required measurable system parameters to the decision maker at every time slot, which only necessitates a few bits for transmitting this information. In addition, the performance of \emph{GMSA} is controlled by the parameter $V$, which achieves the tradeoff between energy cost and average queue backlog. The average queue backlog also reflects the delay performance. Theoretical analysis shows that increasing the value of $V$ enables the achieved time average energy cost to reach arbitrarily near or below the optimal energy cost. Conversely, increasing $V$ will also have the effect of increasing the time average queue backlog, consequently leading to an increase in job processing delay. The analysis shows that the time average energy cost is, at most, $O(1/V)$ above the optimal target, while the time average queue backlog is $O(V)$. Thus, the parameter $V$ can be chosen to push the time average energy cost towards the optimal target with a corresponding tradeoff in time average queue backlog (or delay performance).

\section{Performance Evaluation}\label{section5}

In this section, extensive simulations are conducted to investigate the performance of \emph{GMSA}. First, we introduce the evaluation methodology, including basic configurations, data traces, comparisons, and simulation scenarios. Following that, the performance of \emph{GMSA} is presented and analyzed.

\subsection{Experimental Setup}
\textbf{Basic configurations:} We considers the four data centers of \emph{Facebook}, which are geographically distributed in four regions globally, including Prineville, Oregon \footnote{https://www.facebook.com/PrinevilleDataCenter/}; Forest City, North Carolina \footnote{https://www.facebook.com/ForestCityDataCenter/}; Lulea, Sweden \footnote{https://www.facebook.com/LuleaDataCenter/}; and Altoona, Iowa \footnote{https://www.facebook.com/AltoonaDataCenter/}. One type of data analytics job is simulated in our experiments. The job arrival is based on the production traces from \emph{Facebook}'s Hadoop cluster, that is, $350 K$ jobs per month. Thus, the random job arrival used in our experiments is generated based on the \emph{Poisson} distribution with arrival rate of $350K $ per month. The assumption of \emph{Poisson} based job arrival is reasonable and has been demonstrated by the measurement study \cite{zaharia2010delay}. The weight parameter $w_j(t)$ is set as the real electricity price trace of the four locations, which are obtained from publicly available government agencies. The real-time PUE data traces $PUE_i(t)$ are obtained from the public dashboards of \emph{Facebook}'s data centers. Without loss of generality, the energy consumption from IT component for each job is set as one watt.

To generate the reasonable task allocation ration for each data center, we use the algorithm \emph{Iridium}  in \cite{pu2015low} for minimizing the response time of data analytics. The parameters setting of \emph{Iridium} is similar to the original reference. For instance, each job has fixed $100 GB$ input dataset, which are dynamically distributed in four data centers randomly. The bandwidth values between the data centers and core network vary between 100 Mb/s to 2 Gb/s.

\textbf{Baselines and Metrics:} We compare \emph{GMSA} to two baselines: (i) \emph{DATA}. The fraction of job allocated to any data center is proportional to its dataset distribution. (ii) \emph{RANDOM}. Each job selects its global manager completely random without any consideration. All experiments are conducted based on data sets of 24 hours, and the length of each time slot is set to 5 minutes. Because the purpose of \emph{GMSA} is to minimize the long-term energy cost and balance the tradeoff between cost and delay performance, we collected the time average system cost and queue backlog. The data points provided are averaged from 1000 simulation runs.

\begin{figure*}
  \centering
  \subfigure[Energy cost along time]{
    \label{time_cost} %% label for first subfigure
    \includegraphics[height=6.5cm]{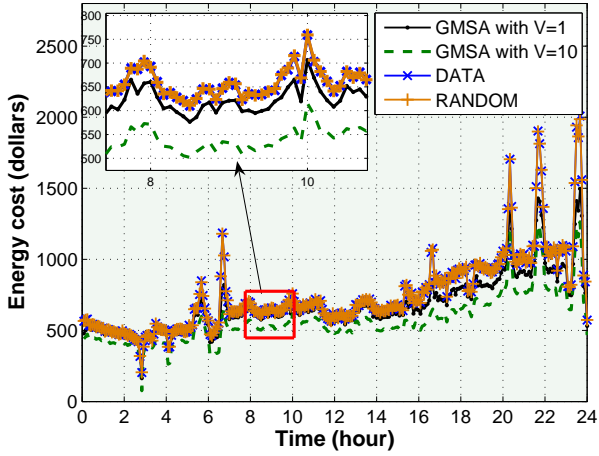}}
  \hspace{-4mm}
  \subfigure[Queue backlogs along time]{
    \label{time_queue} %% label for second subfigure
    \includegraphics[height=6.5cm]{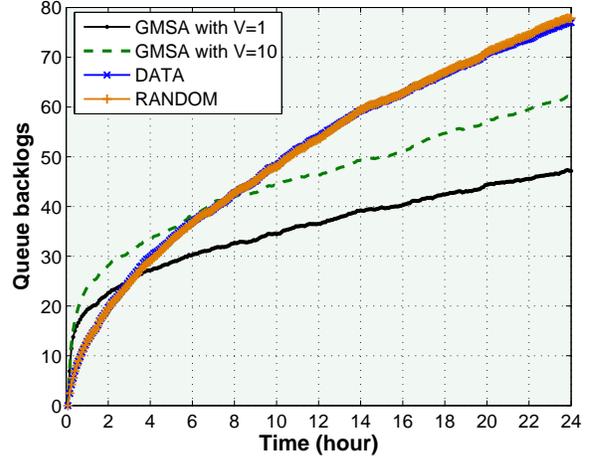}}
  \caption{Performance along time} \label{time_gmsa}
   \vspace{-4mm}
\end{figure*}

\subsection{Evaluation Results}

Two sets of experiments are conducted to investigate the performance of \emph{GMSA} with respect to the performance along time and the sensitivity to control parameter $V$.

\subsection{System performance along time}

Figure~\ref{time_gmsa} illustrates illustrate the performance of \emph{GMSA}, \emph{DATA}, and \emph{RANDOM} over 288 time slots (24 hours). The parameter $V$ of \emph{GMSA} is fixed to be 1 and 10, respectively. From Figure~\ref{time_cost}, we observe that the energy cost of \emph{GMSA} ($V$=1 and $V$=10) is lower than that of both \emph{DATA} and \emph{RANDOM} in almost all time slots. This results verify the effectiveness of the decisions made by \emph{GMSA}. From Figure~\ref{time_queue}, we observe that \emph{GMSA} push the average queue backlogs towards a stable value over time. This results indicate that the dynamic control of \emph{GMSA} could achieve the objective of maintaining system stability. For instance, \emph{GMSA} keeps the average queue backlogs below 50 when $V=1$. The system stability also implies that the time average backlog should not go to infinity, and \emph{GMSA} could provide delay guaranteed service. On the contrary, the time average backlogs of DATA and RANDOM increases dramatically over time and may go to infinity. Overall, these simulation results demonstrate that the dynamic control of GMSA stabilize the system while achieving the optimization objective of minimizing the energy cost.

\subsection{Sensitivity to control parameter $V$}

\begin{figure*}
  \centering
  \subfigure[Energy cost vs. $V$]{
    \label{v_cost} %% label for first subfigure
    \includegraphics[height=6.5cm]{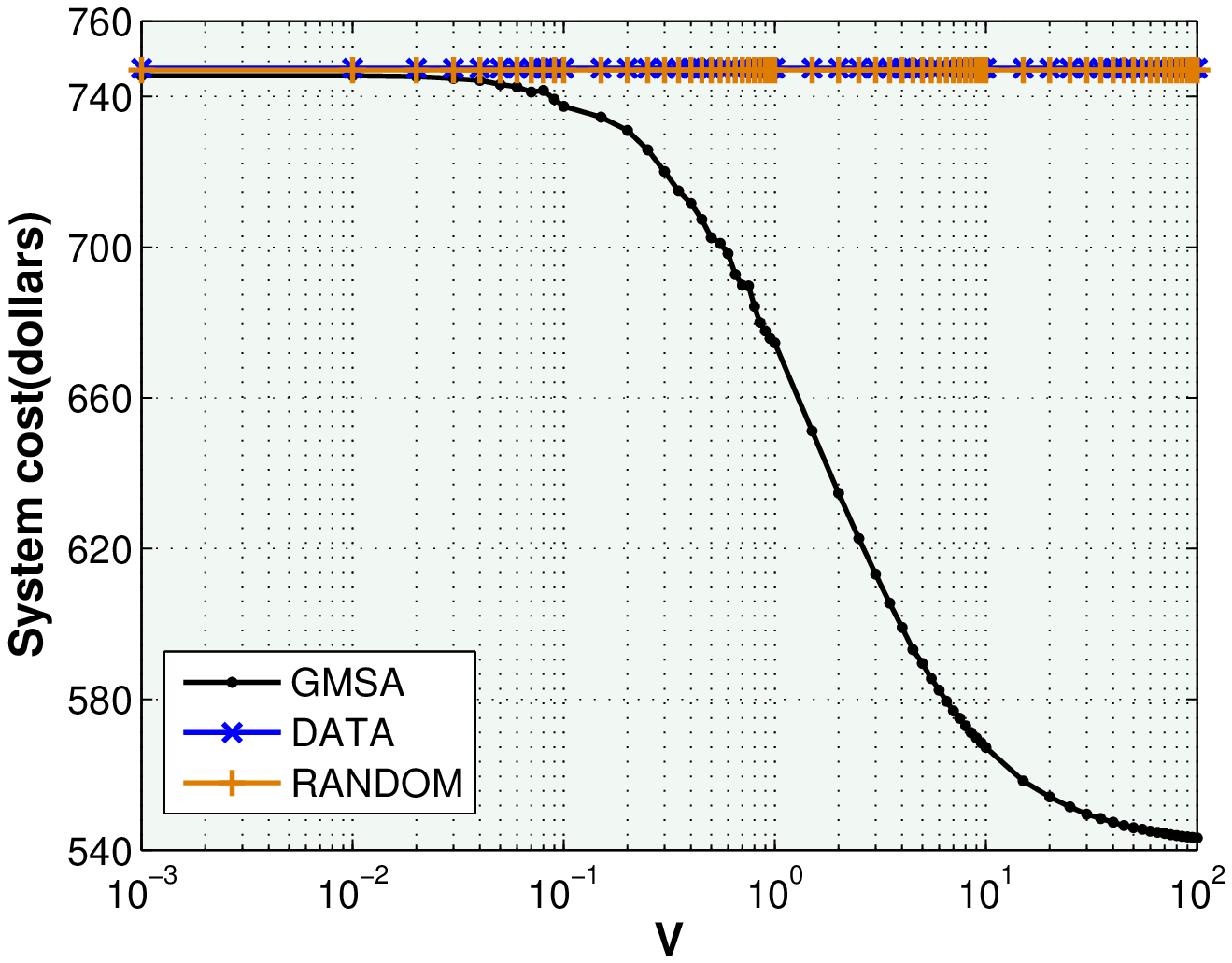}}
  \hspace{-4mm}
  \subfigure[Queue backlogs vs. $K$]{
    \label{v_queue} %% label for second subfigure
    \includegraphics[height=6.5cm]{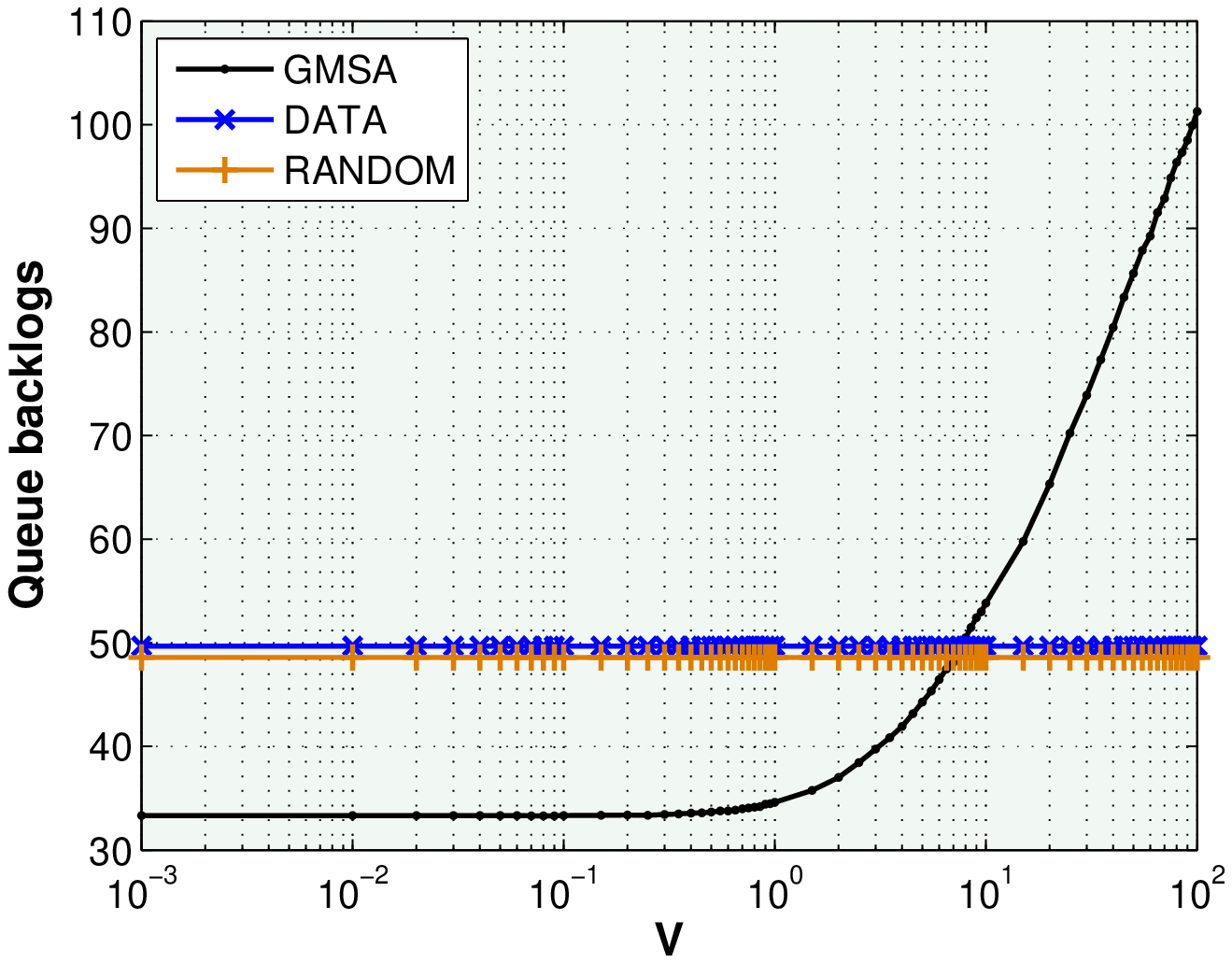}}
  \caption{Sensitivity to control parameter $V$} \label{v_gmsa}
   \vspace{-4mm}
\end{figure*}

Figure~\ref{v_gmsa} illustrates illustrate the performance of \emph{GMSA}, \emph{DATA}, and \emph{RANDOM} by vary $V$ from 0.001 to 100. Because the control parameter $V$ does not affect the performance of \emph{DATA} and \emph{RANDOM}, the corresponding curves in the figure maintain the same value as $V$ increases. We can observe from Figure~\ref{v_cost} that \emph{DATA} and \emph{RANDOM} always present the highest energy cost (approximately 750 dollars), because the manager selection strategies of these two baselines are conducted without considering the cost of energy consumption. On the contrary, our proposed \emph{GMSA} presents the lowest energy cost. Even in the worst case, the three strategies has the similar performance. In the best case, the energy cost of \emph{GMSA} can be as low as 540 dollars. The reduction of energy cost can achieve 30\,\%  approximately. Moreover, we can clearly see that the average energy cost of \emph{GMSA} converges to the minimum value as $V$ increases.

Figure~\ref{v_queue} illustrates the time average queue backlogs of the three strategy by vary $V$ from 0.001 to 100. We can observe that the time average queue backlogs of \emph{GMSA} is much smaller than that of \emph{DATA} and \emph{RANDOM}, and grows to a high level as $V$ increase. When $V$ is larger than 10 approximately, the average queue backlogs of \emph{GMSA} become higher than that of the two baselines. This phenomenon is consist with the theory performance of \emph{GMSA}, that is, increasing the value of $V$ chosen enables the achieved time average system cost being arbitrarily near or below the optimal energy cost. Conversely, increasing $V$ also will have the effect of increasing the time average queue backlog. Overall, the control parameter $V$ can push the time average system cost towards the optimal target with a corresponding tradeoff in time average queue backlog (or delay performance).

\section{Open resource issues}\label{section6}

The research on exploiting the potential benefits of geo-distributed big-data analytics is relatively new and many issues have not been well addressed. Besides the global manager selection discussed in this article, we identify some research opportunities and challenges lie ahead for the design of more efficient GDA system.

\textbf{Placements for Original and Intermediate Data}. The performance of GDA highly depends on the placement of original datasets over different DCs, which can be optimized before the task allocation and execution. In addition, the placement of intermediate data (e.g. the key-value pairs
produceed in \emph{MapRedue}) during the analysis processes also have significantly impact on the GDA performance. Therefore, efficient algorithms for both original and intermediate data placements are highly desired.

\textbf{Task Allocation over DCs}. The global manager is responsible for decomposing each job into multiple tasks and assigning these tasks to different DCs for processing in parallel. Here, task allocation algorithms are essential to achieve load balancing and system efficiency maximization. For instance, if heavy task loads are allocated to a DC with low up and down network I/O bandwidths, network would be congested caused by intensive data transfer, resulting in high response time. Technically, task allocation is challenging, because it needs to jointly consider many system parameters such as original dataset locations, processing efficiency of each individual DC, and network I/O conditions.

\textbf{Fog/Edge computing based GDA.} Currently, most GDA paradigms adopt the server-client architecture with the frontend server and back-end Cloud (i.e. edge-Cloud DCs). However, the long distance interactive across geo-distributed DCs or clusters globally would result in the serious problems of large latency and high cost for big data management and transfer. The emerging data processing paradigm with Fog architecture \cite{shu2017fog} would address these issues by offloading the data processing tasks from the edge-Cloud DCs to Fog servers that are more closer to the users. Many important research issues arise by combining the Fog computing and GDA, including data management and processing over Fog servers and edge-cloud DCs, resource virtualization and allocation, and hybrid Fog and Edge-cloud task decomposition and allocation for GDA jobs.

\section{Conclusion}\label{section7}

In this article, we presented a systemic investigation of energy consumption in big-data analytics over geographically distributed datasets.We first analyzed the specific characterizes of geo-distributed data analysis and its energy consumption patterns. We then presented a dynamic global manager selection algorithm (GMSA) to minimize the energy cost by fully exploiting the system diversities in geography and variation over time. Furthermore, extensive experiments based on the real-world data traces demonstrated the effectiveness and efficiency of our proposed algorithm, reducing energy costs and queue backlogs in comparison with the baseline models. In
addition, we pointed out several potential research directions and challenges that remain open for future studies.

\bibliographystyle{IEEEtran}
\bibliography{gdda_ref}

\end{document}